\documentclass[aps,showpacs,nobibnotes,oneside,onecolumn,floats,eqsecnum,nofootinbib]{revtex4}

\usepackage[unicode=false,draft=false,bookmarks=true]{hyperref}

\usepackage{amsmath,amsfonts,amssymb}

\usepackage{graphicx}
\usepackage{picinpar}

\def\ev#1{\ensuremath{\boldsymbol{#1}}}            
\newcommand{\pder}[2]{\frac{\partial{}#1}{\partial{}#2}} 
\newcommand{\abs}[1]{\ensuremath{\left|{#1}\right|}}     
\newcommand{\pd}{\partial}                         
\newcommand{\Dlr}{\ensuremath{\overleftrightarrow{D}}} 
\newcommand{\avg}[1]{\ensuremath{\left<{#1}\right>}} 
\newcommand{\cket}[1]{\ensuremath{\left|{#1}\right>}}
\newcommand{\brac}[1]{\ensuremath{\left<{#1}\right|}}

\newcommand{\okcol}[2]{\begin{pmatrix}{#1}\\{#2}\end{pmatrix}}

\newcommand{\okmatr}[4]{\begin{pmatrix} {#1}&{#2}\\{#3}&{#4} \end{pmatrix}}

\def\hc{^\dagger}                      
\def\Lagr{\mathcal{L}}                 
\newcommand{\Reals}{\ensuremath{\mathbb{R}}}    

\newcommand{\cross}[2]{[{#1}{#2}]}                   
\newcommand{\comm}[2]{\left[{#1}, {#2}\right]}       

\DeclareMathOperator{\rot}{rot}
\DeclareMathOperator{\divg}{div}
\DeclareMathOperator{\RRe}{\mathtt{Re}}
\newcommand{\ddt}{\frac{\pd}{\pd{}t}}            

\newcommand{\inspic}[3]{
    \begin{figure}[tbh]
      \begin{center}
       \includegraphics[width=#3]{#2.eps}
        \caption{#1}
        \label{pic:#2}
      \end{center}
    \end{figure}
}

{%
  \begin{figwindow}[3,r,
  {\fbox{\includegraphics[width=#3]{#2.eps}}},%
  {#1}\label{pic:#2}]
}%
{\end{figwindow}}

\hypersetup{pdfauthor={O.G.Kharlanov and V.Ch.Zhukovsky, Dep.Theor.Phys., MSU},%
            pdftitle={CPT and Lorentz violation effects in hydrogen-like atoms},%
            pdfsubject={Hydrogen-like atom within SME},%
            pdfcreator={O.K., V.Ch.Zh.},%
            pdfkeywords={Hydrogen,SME,CPT,Lorentz,violation,parity,Dirac,nonrelativistic,
            approximate}}

\newcommand{\ext}{{(\mathrm{e})}}    
\newcommand{\Int}{{\mathrm{int}}}    
\def\SME{\ensuremath{\mathrm{SME}}}  
\def\CPT{\ensuremath{\mathrm{CPT}}}  
\def\Parity{\ensuremath{\mathrm{P}}} 

\begin{document}
\title{CPT and Lorentz violation effects in hydrogen-like atoms
       \footnote{The article is published in J. Math. Phys. \textbf{48}, Issue 9, p.092302 (September 2007).}
       }
\author{O.~G.~Kharlanov}\email{okharl@mail.ru}
\author{V.~Ch.~Zhukovsky}\email{zhukovsk@phys.msu.ru}
\affiliation{Department of Theoretical Physics,
             Moscow State University, 119992 Moscow, Russia}
\pacs{12.60.-i, 32.10.-f, 03.65.Ge, 31.30.Jv, 32.30.-r} 
\begin{abstract}
    Within the framework of Lorentz-violating extended electrodynamics, the Dirac equation
    for a bound electron in an external electromagnetic field is considered assuming the interaction
    with a \CPT-odd axial vector background $b_\mu$.  The quasi-relativistic Hamiltonian
    is obtained using a $1/c$-series expansion. Relativistic Dirac eigenstates in a
    spherically-symmetric potential are found accurate up to the second order in $b_0$.
    $b_0$-induced \CPT-odd corrections to the electromagnetic dipole moment operators
    of a bound electron are calculated that contribute to the anapole moment of the atomic
    orbital and may cause a specific asymmetry of the angular distribution of the
    radiation of a hydrogen atom.
\end{abstract}
\maketitle
\section{Introduction}\label{sec1}
The Standard Model is currently proved with a convincingly wide set of experiments.
Nevertheless, it essentially does not include a quantum description of gravitation. The
quantization methods adopted in the Standard Model do not allow a self-consistent
quantization of General Relativity since the subsequent theory occurs to be
nonrenormalizable. Thus the essence of the theory accounting for the effects taking place
at the \textit{Planck scale} of energies ($E_{\text{Pl}}=10^{19}$GeV) where quantum
gravity plays a major role, still remains obscure. At the same time, there exist some
candidates for such Fundamental theory, string theory for instance, taking the form of
the Standard Model in the low-energy limit.

Planck energies being far from experimental attainment, the \textit{Standard Model
Extension} (\SME) was elaborated. It is an effective theory (applicable at the energies
$E \ll E_{\text{Pl}}$) formulated axiomatically as a set of corrections to the Lagrangian
of the Standard Model fulfilling some `natural' requirements \cite{ColladayVAK:SME,
Bluhm:SME} such as observer Lorentz invariance, 4-momentum conservation, unitarity, and
microcausality. In what follows, we will focus on a subset of the \SME{} referred to as
the \textit{minimal} \SME{} in \textit{flat} Minkowsky spacetime that also requires local
$SU(3)_C\times SU(2)_I\times U(1)_Y$ gauge invariance and power-counting
renormalizability. A spectacular feature of such requirements is that they reduce the
diversity of possible corrections down to a \textit{finite} number of them. Each
correction term consists of a complex (pseudo)tensor constant (\SME{} coefficient)
contracted with conventional Standard Model fields and their spacetime derivatives. These
constants are believed to stand for vacuum expectation values of the fields featuring in
the hypothetic Lorentz-covariant Fundamental theory and condensed at low energies due to
the spontaneous symmetry breaking mechanism. Indeed, it has been shown recently that such
Lorentz symmetry breaking can occur in some theories beyond the Standard Model
\cite{SpontLCPTBreaking:1, SpontLCPTBreaking:2, SpontLCPTBreaking:3, SpontLCPTBreaking:4}
leading subsequently to the \SME{}. The \SME{} can thus be used to reduce the complexity
of these theories and related calculations in the low-energy limit. It also provides a
standard for representation of data obtained in experiments searching for Lorentz
violation.

Recently, a number of theoretical researches have been performed aiming at investigating
the vacuum structure of this model (see, e.g. \cite{VAKLehnert, EbZhRazum, Andrianov,
Altschul:PV, Jackiw}), and to study the assumed violation on various high-energy
processes \cite{VAKPickering, AdamKlink, Altschul:SR, Lena}. This search also seems quite
promising in atomic physics \cite{Bluhm:Atom, BluhmVAKRussell}. For instance, specific
types of Lorentz violation may cause spatial parity violation in electrodynamics at tree
level. \Parity-parity violation effects in atomic systems \textit{within the conventional
Standard Model} have been thoroughly studied either theoretically or experimentally in
the past four decades \cite{Zeldovich:Anapole, CurtisMichel:WeakCurrents,
Bouchiat:WeakCurrents, Khriplovich:AtomicParity2, Khriplovich:AtomicParity3}. Such
effects are caused by weak interaction and include resonant dichroism of atomic gas,
permission of conventionally forbidden quantum transitions etc. Much the same effects are
expected within the \SME{}.

Until today however, studies of atom within the \SME{} have included only spectroscopic
predictions using the perturbation theory with respect to \SME{} coefficients
\cite{Bluhm:Atom, BluhmVAKRussell, Hydrogen1, Hydrogen3}. Direct solution of the atomic
eigenstate problem would make it possible to study \textit{radiative} properties of the
atom, too. This paper is devoted to an analysis of dynamics of a charged fermion in an
external electromagnetic field within extended electrodynamics with a background axial
vector $b_\mu$. Approximate methods are used to implicitly solve the eigenstate problem
in a central electric field; for the Coulomb field, an explicit solution is obtained (see
section \ref{sec5}). The quasi-relativistic approach is also employed to obtain the
corrections the conventional Schr\"{o}dinger-Pauli-Dirac equation acquires in the
background of $b_0$ (sections \ref{sec3}, \ref{sec4}). Moreover, in view of recent
publications \cite{VAKLane:Hamiltonian, Brazil:SMEDirac, Belich}, some other types of
couplings and the corresponding fermion eigenstates are also discussed. As an example of
applying these results, an effect of $b_0$-induced asymmetry of the angular distribution
of spontaneous radiation of a polarized hydrogen atom is demonstrated (section
\ref{sec6}).

In addition, a polarized hydrogen atom is shown to have a nonzero \textit{anapole} moment
originally introduced in \cite{Zeldovich:Anapole}. This characteristic is specific for
parity-nonconserving systems interacting with electromagnetic field. For example, due to
weak interaction loops, neutrinos can possess such a moment, and it is the only
electromagnetic characteristic that maintains for Majorana neutrinos
\cite{BorZhukTern:Anapole}.
\section{The Model}\label{sec2}
We will restrict our consideration to a specific case of extended electrodynamics of electrons and photons
within the \SME{} (further refereed to as extended QED) with the Lagrangian
\begin{eqnarray}
   \Lagr &=& -\frac14 F_{\mu\nu}F^{\mu\nu}
        +\bar\psi\left(\frac{i}{2}\gamma^\mu \Dlr_\mu - m_e - b_\mu\gamma^\mu\gamma_5\right)\psi,
                   \label{L}\\
   D_\mu &\equiv& \pd_\mu + i e A_\mu(x),
   \qquad \gamma_5 = -i \gamma^0\gamma^1\gamma^2\gamma^3,
\end{eqnarray}
where $e$, $m_e$ are the electron charge and mass, respectively;
$x^\mu\equiv\{ct,\ev{r}\}$, and $b_\mu$ is a constant \CPT-odd axial vector. Present
constraints on $b_\mu$ \textit{for electron} are the following \cite{Bluhm:Atom,
BluhmVAKRussell, Coleman}:
\begin{eqnarray}
  \mid b_0 \mid  &\lesssim& 10^{-2}\mbox{ eV},\\
  \mid\ev{b}\mid &\lesssim& 10^{-19}\mbox{ eV},
\end{eqnarray}
while the constraints for \textit{nucleons} are some orders more stringent. We will use
one-particle approximation in the framework of relativistic quantum mechanics. Recalling
\eqref{L}, one can obtain the Hamiltonian for an electron in an external electromagnetic
field
\begin{equation}\label{H}
      \hat{H}(t) = \ev\alpha\cdot\hat{\ev{P}} +
                \beta m_e + e A_0(\hat{\ev{r}},t) +
                b_0\gamma_5 + \ev{b} \cdot \ev\Sigma,
\end{equation}
where $\hat{\ev{P}} \equiv \hat{\ev{p}}-e\ev{A}(\hat{\ev{r}},t)$ and $\ev\alpha \equiv
\gamma^0\ev\gamma$, $\beta \equiv \gamma^0$, and $\ev\Sigma \equiv -\ev\alpha\gamma_5$.
In Dirac spinor representation, spatial parity operator takes the form:
\begin{equation}
   \hat{P}\xi(\ev{r}, t) \equiv \gamma^0\xi(-\ev{r}, t),
\end{equation}
hence $\hat{P}\hc = \hat{P}$, $\hat{P}\hc\hat{P} = \hat{P}^2 = 1$. Hamiltonian \eqref{H}
commutes with $\hat{P}$ if  $A_0 (\ev{r},t) = A_0(-\ev{r},t)$, $\ev{A}(\ev{r},t) =
-\ev{A}(-\ev{r},t)$, and  $b_0 = 0$. In particular, the presence of $b_0$ can violate the
\Parity-parity of Hamiltonian \eqref{H} in a spherically-symmetric field $A^\mu =
\{\phi(r), \ev{0}\}$, in the Coulomb field of an infinitely heavy nucleus for example,
with
\begin{equation}
    \phi(r) = -\frac{Z e}{4\pi r}.
\end{equation}
\textit{Charge} conjugation of Hamiltonian \eqref{H} only changes the sign of electric
charge $e$ to the opposite; so hydrogen and anti-hydrogen atoms possess equivalent
dynamics even if $b_\mu \ne 0$. Due to these facts our primary interest concerns the
investigation of unusual properties of a hydrogen atom induced by the presence of a
nonzero constant $b_0$.
\section{$1/c^2$-approximation for the Dirac equation in the $b_\mu \ne 0$ case}\label{sec3}
The quasi-relativistic approximation assumes an expansion into a
series with respect to $1/c$. It is thus necessary in this section
to rewrite the Hamiltonian \eqref{H} in the \textrm{CGS} system of
units with the speed of light $c \ne 1$
\begin{equation}\label{eq:H_D}
      \hat{H} = c\ev\alpha\cdot\hat{\ev{P}} +
                \beta m_e c^2 + e A_0 + c b_t\gamma_5 + \ev{b} \cdot \ev\Sigma,
\end{equation}
where $\hat{\ev{P}} \equiv
\hat{\ev{p}}-\frac{e}{c}\ev{A}(\hat{\ev{r}},t)$ and $b_t \equiv
b_0/c$ so that $b_t$ has dimensionality of momentum. Consider the
extended Dirac equation with the Hamiltonian \eqref{eq:H_D} in a
non-stationary external field $A_\mu(x)$:
\begin{gather} \label{eq:Dirac}
    i\hbar\pder{\psi(\ev{r}, t)}{t}=\hat{H}(t)\psi(\ev{r},t),\\
    \int\psi\hc(\ev{r},t)\psi(\ev{r},t)d^3r = 1.
\end{gather}
Following the standard method (see, e.g. \cite{LL4, STZ}), let us shift the energy by
means of a unitary transformation
\begin{equation}\label{e24}
    \psi = \exp\left\{-i \frac{m_ec^2}{\hbar} t\right\} \okcol{u}{v}.
\end{equation}
In terms of 2-component spinors $u$ and $v$, the Dirac equation \eqref{eq:Dirac} takes
the form:
\begin{gather}
    \label{e25}
    \okmatr{\hat\lambda}{c\hat\Lambda}{c\hat\Lambda}{\hat\lambda - 2m_ec^2}%
    \okcol{u}{v} = 0,\\
    \label{eq:lL}
    \hat\Lambda \equiv \ev\sigma\cdot\hat{\ev{P}} - b_t, \quad
    \hat\lambda \equiv e A_0 + \ev\sigma\cdot\ev{b} - i\hbar\ddt,
\end{gather}
where $\ev\sigma$ denotes a vector of the three Pauli matrices, and
the Dirac matrices are taken in the standard representation:
\begin{equation}
    \beta \equiv \gamma^0 = \okmatr{1}{0}{0}{-1},\quad
    \ev\gamma = \okmatr{\ev{0}}{\ev\sigma}{-\ev\sigma}{\ev{0}},\quad
    \ev\alpha\equiv\gamma^0\ev\gamma =\okmatr{\ev{0}}{\ev\sigma}{\ev\sigma}{\ev{0}},\quad
    \ev\Sigma = \okmatr{\ev\sigma}{\ev0}{\ev0}{\ev\sigma}.
\end{equation}
Consider an electron in a state with a positive sign of energy (this in fact does not
imply that the electron possesses a \textit{definite energy}). External fields are
assumed to be weak enough and to have frequencies much smaller than $m_ec^2/\hbar$ so
that $\ev{E}, \ev{H}, \hat{\ev{P}}, \, i\hbar\ddt - e A_0 = O(c^0)$, when acting upon
$u$, $v$.

In this section, we also assume that
\begin{equation}\label{eq:b_orders}
b_t \equiv b_0/c = O(c^0), \qquad \ev{b} = O(c^0).
\end{equation}
Contrary to the conventional electrodynamics, within the context of which the Gaussian units are usually used,
the order of $b_\mu$ in $1/c$ is quite ambiguous. In electrodynamics, certain powers of $1/c$ can be assigned to
the fields $\ev{E}, \ev{H}, A_\mu$ which result in a hierarchy of electromagnetic effects having different
orders in $1/c$. For instance, radiative processes are at least of the third order in $1/c$, hence the
$1/c^2$-approximation is worth considering. In contrast, the physical origin of $b_\mu$ is not yet finally
established, and hence, we use the convention \eqref{eq:b_orders}, due to the symmetry between
$\ev\sigma\cdot\hat{\ev{P}}$ and $b_t$, both entering \eqref{e25}. Together with \eqref{eq:lL}, this convention
implies
\begin{equation}\label{eq:lL'}
    \hat{\lambda}, \hat{\Lambda} = O(c^0),
\end{equation}
when acting upon $u$, $v$. Then the second line of \eqref{e25} gives:
\begin{equation} \label{e26}
    v = \frac{1}{2m_ec}\left(1+\frac{\hat\lambda}{2m_ec^2}\right)\hat\Lambda u  +
    O(1/c^4),
\end{equation}
and $v$ is thus suppressed, compared with $u$, for nonrelativistic positive-energy
solutions. On the other hand, the square of the norm
\begin{eqnarray}
  \|\psi\|^2 &\equiv& \int{\psi\hc\psi}d^3r = 1 = O(c^0),\\
  \|\psi\|^2 &=& \int(u\hc u + v\hc v)d^3r = \int{u\hc(1+O(1/c^2))u}d^3r,
\end{eqnarray}
consequently, $u = O(c^0)$ and, due to \eqref{e26}, $v = O(1/c)$. This, in addition, results in suppression of
the terms in the quasi-relativistic Hamiltonian stemming from the block-off-diagonal part of the matrix in
\eqref{e25}, in particular, the terms containing $b_0$. As a result, the quasi-relativistic Hamiltonian will
contain leading-order $b_0$-induced contributions proportional to $b_t = b_0/c$, but not $b_0$ itself (see
\eqref{eq:HQuasiRel}).

Now, instead of $u$, the following 2-component spinor field $\Phi(x)\in\mathbb{C}^2$ should be introduced as the
quasi-relativistic wavefunction of the electron:
\begin{equation}
  \Phi(x) \equiv \left(1+\frac{\hat\Lambda^2}{8m_e^2c^2}\right)u. \label{e27}
\end{equation}
In this case, integration by parts shows the time evolution to preserve the norm
\begin{eqnarray}
    \|\Phi\|^2
    \equiv\int{\Phi\hc\Phi}d^3r
    &=&\int{d^3r~\left[\left(1+\frac{\hat\Lambda^2}{8m_e^2c^2}\right)u\right]\hc
            \left[\left(1+\frac{\hat\Lambda^2}{8m_e^2c^2}\right)u\right]}=\nonumber\\
    &=&\int{d^3r~u\hc\left(1+\frac{\hat\Lambda^2}{8m_e^2c^2}\right)^2u}
     = \int{d^3r~u\hc\left(1+\frac{\hat\Lambda^2}{4m_e^2c^2}\right)u} + O(1/c^3)
                                                =\nonumber\\
    &=&\int{d^3r~\left\{u\hc u + \left(\frac{\hat{\Lambda}u}{2m_ec}\right)\hc
       \left(\frac{\hat\Lambda u}{2m_ec}\right) \right\}} + O(1/c^3) =\nonumber\\
    &=&\int{d^3r~\left\{u\hc u + v\hc v\right\}} + O(1/c^3)
    = \|\psi\|^2 + O(1/c^3)
    = 1 + O(1/c^3),
\end{eqnarray}
while $\int{u\hc u}d^3r$ varies with time by a $O(1/c^2)$-amount. However, the
transformation \eqref{e27} leaves ``probability distribution'' $\Phi\hc\Phi$ different
from $\psi\hc\psi$ by a fully-divergent term of the order $1/c^2$:
\begin{gather}
    \label{e22}
    \Phi\hc\Phi = \psi\hc\psi +
    \divg\ev{\mathfrak{j}}_{\text{ZB}} + O(1/c^3),\\
    (\mathfrak{j}_{\text{ZB}})_i =
             -\frac{1}{8m_e^2c^2}\left(\hbar^2\nabla_i(u\hc u)
             -2\hbar\epsilon_{ijk} u\hc \sigma_j \hat{P}_k u\right).
\end{gather}
This situation reflects the presence of negative-energy states resulting in the
\textit{Zitterbewegung} of the electron.
\par Now expressing $u$ and $v$ in terms of $\Phi$ using \eqref{e26} and \eqref{e27},
write the first line of \eqref{e25}:
\begin{equation}\label{e28}
    0 = \hat\lambda u + c \hat\Lambda v =
    \left\{\hat\lambda + \frac{1}{2m_e}\hat\Lambda
         \left(1+\frac{\hat\lambda}{2m_ec^2}\right)\hat\Lambda\right\}
         \left(1- \frac{\hat\Lambda^2}{8m_e^2c^2}\right)\Phi +
         O(1/c^3).
\end{equation}
To obtain an equation in the form $i\hbar\hspace{2pt}\pd\Phi/\pd{t} = \hat{h}\Phi$, one
must make iterations to leave only one time derivative of $\Phi$ in the right side of
\eqref{e28}. The corresponding operator is implicitly contained in $\hat\lambda$. First
consider the above equation in the $1/c$-approximation:
\begin{equation}\label{e28'}
  \hat\lambda\Phi = -\frac{\hat\Lambda^2}{2m_e}\Phi + O(1/c^2).
\end{equation}
After some transformations with the use of \eqref{e28'}, we obtain in the
$1/c^2$-approximation:
\begin{equation}
      \label{eq:GeneralQuasiDirac}
      \left\{ \hat\lambda
             + \frac{\hat\Lambda^2}{2m_e}\left(1-\frac{\hat\Lambda^2}{4m_e^2c^2}\right)
             - \frac{1}{8m_e^2c^2}\comm{\comm{\hat\lambda}{\hat\Lambda}}{\hat\Lambda}
             \right\}\Phi = O(1/c^3).
\end{equation}
Note that commutator $\comm{\hat{\lambda}}{\hat\Lambda}$ does not contain $\pd/\pd{t}$
operator, so there is \textit{only one} time derivative of $\Phi$ in
\eqref{eq:GeneralQuasiDirac}, namely the one contained in $\hat\lambda\Phi$. Converted
into its usual form, \eqref{eq:GeneralQuasiDirac} gives the quasi-relativistic equation
for a positive-energy electron
\begin{eqnarray}\label{e1:quasiDirac}
    i\hbar\pder{\Phi}{t} &=& \hat{h}\Phi, \qquad \psi\hc\psi = \Phi\hc\Phi
    - \divg\ev{\mathfrak{j}}_{\text{ZB}} + O(1/c^3);\\
    \hat{h} &=& \frac{\hat{\Pi}'^2}{2m_e}
    \left(1-\frac{\hat{\Pi}'^2}{4m_e^2c^2}\right)
    -\frac{e\hbar}{2m_ec}\ev{\sigma}\ev{H} + \ev{\sigma}\ev{b}+ eA_0-  \nonumber\\
    &-&\frac{e\hbar}{4m_e^2c^2}\ev{\sigma}[\ev{E}\hat{\ev{P}}]
    -\frac{e\hbar^2}{8m_e^2c^2}\divg\ev{E}
    -\frac{\ev{\sigma}[\hat{\ev{P}}[\ev{b}\hat{\ev{P}}]]}{2m_e^2c^2}, \label{eq:HQuasiRel}\\
    \hat{\ev{\Pi}} &\equiv& \hat{\ev{P}} - b_t\ev{\sigma},\\
    \hat{\Pi}'^2 &\equiv& \hat{\ev{\Pi}}^2 - 2b_t^2 \equiv
                 \hat{\ev{P}}^2 + b_t^2 - 2 b_t
                 \ev{\sigma}\cdot\hat{\ev{P}}.
\end{eqnarray}
The Hamiltonian $\hat{h}$ is precisely hermitian and the corresponding equations of
motion demonstrate their exact local gauge invariance:
\begin{eqnarray}
    \hat{h}\hc[A_\mu]&=&\hat{h}[A_\mu],\\
    \left(\hat{h}[A_\mu]-i\hbar\ddt\right)\exp\left\{i \frac{e}{\hbar c}
    \alpha(x)\right\}&=&
    \exp\left\{i \frac{e}{\hbar c} \alpha(x)\right\}
             \left(\hat{h}[A_\mu + \pd_\mu\alpha]-i\hbar\ddt\right) \qquad
             \forall\alpha(x)\in\Reals.
\end{eqnarray}
\par In the $1/c$-approximation, we arrive at the Pauli equation, through
which the expressions for the probability current and density are easily found:
\begin{eqnarray}
     i\hbar\pder{\Phi_P}{t} &=& \hat{h}_P\Phi_P, \label{eq:PauliTimeEvolution}\\
     \label{eq:hPauli2}
      \hat{h}_P  &=& \frac{\ev\Pi^2}{2m_e} - \frac{b_t^2}{m_e}
                 - \frac{e\hbar}{2m_ec}\ev{\sigma}\ev{H} + eA_0
                 +  \ev{\sigma}\ev{b},\\
     j^{\mu}_P &=& \left\{ c \Phi_P\hc\Phi_P,\quad \frac{1}{2 m_e}\left(
                \Phi_P\hc (\hat{\ev{P}} \Phi_P)
                + (\hat{\ev{P}} \Phi_P)\hc\Phi_P\right) - \frac{b_0}{m_e c}
                \Phi_P\hc\ev{\sigma}\Phi_P \right\},
\end{eqnarray}
i.e. the current acquires an additional spin-dependent term in the $b_0 \ne 0$ case. The
terms in \eqref{eq:hPauli2} involving external fields form the interaction Hamiltonian.
For $A_\mu(x)$ taken in the Coulomb gauge, it reads as follows:
\begin{eqnarray}\label{eq:PauliInt}
    \hat{h}_{P\,\Int} &=& -\frac{e}{m_e c}\ev{A}\cdot\hat{\ev{\pi}} + e A_0
                      - \frac{e\hbar}{2m_ec}\ev\sigma\cdot\ev{H}
                      + \frac{e^2}{2m_ec^2}\ev{A}^2,\\
    \hat{\ev{\pi}} &\equiv& \hat{\ev{p}} - b_t \ev\sigma.
\end{eqnarray}
The difference of \eqref{eq:PauliInt} from that in the conventional QED is generated by a
gauge-like shift in the momentum space ($\hat{\ev{p}} \to \hat{\ev\pi}$). This feature
will be used in section \ref{sec4} for constructing the solutions of the eigenstate
problem.
\par The results obtained agree with those published in \cite{VAKLane:Hamiltonian}
and \cite{Brazil:SMEDirac}, in the corresponding particular cases. In the former paper, a
nonrelativistic Hamiltonian for a \textit{free} electron was obtained using the
Foldy-Wouthysen method ($1/m_e$-series), within the first order approximation with
respect to \textit{all possible} \SME{}-corrections in the fermion sector of extended QED
\cite{ColladayVAK:SME, Bluhm:SME}. In the special case of the axial vector background
$b_\mu$, the resulting nonrelativistic Hamiltonian can be obtained from the formulas of
paper \cite{VAKLane:Hamiltonian}:
\begin{equation}
  \label{eq:h_rel}
  \hat{h}_{\textrm{FW}} = \frac{\hat{\ev{p}}^2}{2m_e}
  + \ev\sigma\ev{b} - \frac{b_0}{m_ec}\ev{\sigma}\hat{\ev{p}}
  + \frac{\hat{p}_j \sigma_l}{2m_e^2c^2}(b_j \hat{p}\,_l - b_l \hat{p}_j)
  + \frac{b_0}{2m_e^3c^3}\hat{\ev{p}}^2(\ev\sigma\hat{\ev{p}}).
\end{equation}
On the other hand, for a free electron, the Hamiltonian \eqref{eq:HQuasiRel} takes the
form:
\begin{equation}
   \hat{h} = \frac{\hat{\ev{p}}^2 - 2b_t \ev\sigma\cdot\hat{\ev{p}} + b_t^2}{2m_e}
    \left(1-\frac{\hat{\ev{p}}^2 - 2b_t \ev\sigma\cdot\hat{\ev{p}}+b_t^2}{4m_e^2c^2}\right)
    +\ev{\sigma}\ev{b}
    -\frac{\ev{\sigma}[\hat{\ev{p}}[\ev{b}\hat{\ev{p}}]]}{2m_e^2c^2}.
\end{equation}
One can easily find that, within the linear order in $b_\mu$ and the
third order in $p/m_e$ (the approximation used in
\cite{VAKLane:Hamiltonian}), the two expressions are identical. The
absence of the term proportional to $p^4/m_e^3$ in the former
expression does not indicate an error. Instead, it is a consequence
of the difference in the expansion parameters chosen, i.e. $p/m_e$
and $1/c$, respectively.

It should be emphasized that the method used in paper
\cite{VAKLane:Hamiltonian} to obtain expression \eqref{eq:h_rel} was
based on a series expansion of a \textit{precise relativistic}
Hamiltonian \textit{for a 2-component wavefunction of a free
particle} constructed using the Foldy-Wouthysen iterations
\cite{FW}. Making these iterations, however, is inconvenient in the
presence of external fields. In contrast, the method used in our
paper takes these fields into account from the beginning.

Quasi-relativistic methods similar to those used in our paper were employed in
\cite{Brazil:SMEDirac} to find the $1/c$-corrections to the Dirac equation \textit{in an
external electromagnetic field} with additional $a_\mu$ and $b_\mu$ \SME{}-couplings. In
addition, plane wave solutions were obtained, and $\SME$-specific modifications of the
hydrogen spectrum were estimated, within the nonrelativistic approximation. For the
$b_\mu$ coupling, the calculations performed have led to the Pauli Hamiltonian of the
form \eqref{eq:hPauli2}.
The contributions in the fermion Lagrangian and Hamiltonian, corresponding to $a_\mu$
coupling, are as follows \cite{ColladayVAK:SME, Bluhm:SME}:
\begin{eqnarray}
   \Delta\Lagr^{(a)} &=& - \bar\psi \gamma^\mu a_\mu \psi,\\
   \Delta\hat{H}^{(a)} &=& \gamma^0 \gamma^\mu a_\mu = a_0 - \ev\alpha\cdot\ev{a},
\end{eqnarray}
where $a_\mu$ is a constant background 4-vector, which can be treated as a vacuum
expectation value of some Planck-scale fundamental fields. As mentioned in
\cite{ColladayVAK:SME, Brazil:SMEDirac},  transition from the $a_\mu = 0$ to the $a_\mu
\ne 0$ case is a kind of a gauge transformation because
\begin{eqnarray}
    A_\mu(x) &\to& A_\mu^{(a)}(x) = A_\mu(x) + \frac{1}{e}a_\mu
        = A_\mu(x) - \pd_\mu\alpha(x), \label{e32a}\\
    \alpha(x) &=& -\frac{1}{e}a_\mu x^\mu.
\end{eqnarray}
This feature makes it possible to find a system of exact solutions of the Dirac equation modified with the
$a_\mu$-term making an inverse gauge (and unitary) transformation. Suppose the eigenstate problem is solved in
the $a_\mu = 0$ case so that
\begin{gather}
    \hat{H}^{(0)}\psi_{n}^{(0)} = E_{n}^{(0)} \psi_{n}^{(0)},\\
    \hat{n}_i^{(0)}\psi_{n}^{(0)} = n_i\psi_n^{(0)}, \quad i = 1,2,\ldots,N; \label{eq:n_i^(0)}\\
    \left(\psi^{(0)}_{m}, \psi^{(0)}_{n}\right) \equiv
    \int{d^3r}{\ \psi_{m}^{(0)\dagger}(\ev{r})\psi_{n}^{(0)}(\ev{r})} = \delta_{m,n},
\end{gather}
where $m = \{m_i\} \equiv \{m_1, m_2,\ldots, m_N\}$, $n = \{n_i\}\equiv \{n_1, n_2, \ldots, n_N\}$ denote the
sets of quantum numbers corresponding to $N$ hermitian operators $\hat{n}_i^{(0)}$ that should commute with
$\hat{H}^{(0)}$ and with each other. The operators $\hat{n}_i^{(0)}$ are needed only to represent quantum
numbers, i.e. they form a \textit{complete set of observables}. It should be pointed out that the choice of
these operators does not affect the eigenstate problem itself, but only forms the basis of the eigenstates and
enumerates them. For example, for the nonrelativistic hydrogen atom the quantum numbers are usually taken such
that $n_1 \equiv n$, $n_2 \equiv l$, $n_3 \equiv m$ define the eigenvalues of the three operators, namely
$\hat{n}_1^{(0)} \equiv \hat{H}^{(0)}$, $\hat{n}_2^{(0)} \equiv \hat{\ev{l}}^2$, and $\hat{n}_3^{(0)} \equiv
\hat{l}_3$, in the eigenstate $\psi^{(0)}_{n l m}$.

The system of solutions for $a_\mu \ne 0$ reads
\begin{gather}
    \hat{H}^{(a)}\psi_{n}^{(a)}(\ev{r}) = E_{n}^{(a)} \psi_{n}^{(a)}(\ev{r}),\\
    \hat{n}_i^{(a)} \psi^{(a)}_n = n_i \psi^{(a)}_n,  \label{e35a}\\
    \left(\psi^{(a)}_{m}, \psi^{(a)}_{n}\right) = \delta_{m,n};\\
    \hat{H}^{(a)} = c\ev\alpha\left(\hat{\ev{p}} - \frac{1}{c}(e\ev{A}+\ev{a})\right)
                    + m_e\beta + (eA_0+a_0);\\
    \psi^{(a)}_{n} = e^{i\ev{a}\cdot\ev{r}/\hbar c}\psi^{(0)}_n, \label{e33} \\
    E_n^{(a)} = E^{(0)}_n + a_0. \label{e34}
\end{gather}
The energy spectrum is shifted by a constant value $a_0$; no spectroscopic signature is therefore left by the
presence of the nonzero $a_\mu$ (i.e. transition frequencies are unaffected). However, the meaning of the
quantum numbers $n_i$ (which run through the same set of values as in \eqref{eq:n_i^(0)}) is changed, because,
for $\psi^{(a)}_n$ functions, they correspond to the operators $\hat{n}_i^{(a)} \ne \hat{n}^{(0)}_i$ which can
be readily constructed from $\hat{n}_i^{(0)}$:
\begin{equation}
    \hat{n}_i^{(a)} = e^{i\ev{a}\hat{\ev{r}}/\hbar c}\hat{n}^{(0)}_i
                       e^{-i\ev{a}\hat{\ev{r}}/\hbar c}. \label{e35b}
\end{equation}
 For instance, in the case of a hydrogen atom, when the unitary transformation is made,
$\psi^{(0)}_{nlm}\to\psi^{(a)}_{nlm}$, with $a$ being some parameter of the
transformation, the resulting $\psi^{(a)}_{nlm}$ is an eigenstate of the transformed
Hamiltonian $H^{(a)}$, with the same eigenvalue (energy), but now quantum numbers $n,l,m$
correspond to \textit{new} operators
$\hat{n}_i^{(a)}\ne\hat{n}_i^{(0)}$, $i = 1,2,3$.

Using \eqref{e27} and \eqref{e24}, one can find that the transformation \eqref{e33} maintains its form for the
Pauli wavefunction $\Phi_P$:
\begin{eqnarray}
  \Phi^{(a)}_{P,n}(\ev{r}) &=& e^{i\ev{a}\cdot\ev{r}}\Phi_{P,n}(\ev{r}),
\end{eqnarray}
while the nonrelativistic spectrum and $\hat{n}_i$ operators are still transformed
following \eqref{e34} and \eqref{e35b}, respectively.

Another investigation, which is worth mentioning, was held in \cite{Belich}. The authors
have considered two non-minimal Lorentz-violating couplings in the fermion sector of QED:
\begin{equation}
  \Delta\Lagr^{(g,g_{\mathrm{a}})} = \bar\psi(-g v^\nu + g_{\textrm{a}} v_{\textrm{a}}^\nu
                                 \gamma_5) \gamma^\mu F^*_{\mu\nu}\psi,
\end{equation}
where $g$ and $g_{\textrm{a}}$ are the coupling constants while $v^\nu$ and
$v_{\textrm{a}}^\nu$ are fixed background vectors (`$\textrm{a}$' is not a component
index but means `axial'), and $F^*_{\mu\nu} =\frac12\epsilon_{\mu\nu\alpha\beta}
F^{\alpha\beta}$ is the dual field tensor. We leave aside the question of the origin of
such couplings; for more information, the reader is referred to \cite{Belich} and the
references therein. The authors of the paper also used the nonrelativistic Pauli approach
to obtain the Pauli equation in the presence of the background vectors $v^\nu$ and
$v_{\textrm{a}}^\nu$, and then calculated the first-order energy corrections using
perturbation theory. In addition, they considered an atom in a homogeneous external
magnetic field also treated perturbatively. The case of the $g$ and $g_{\textrm{a}}$
couplings, in general, does not permit such an easy construction of the eigenfunctions as
the case of the $a_\mu$ coupling. However, in one special case not considered in
\cite{Belich}, namely for the constant homogeneous external field $F_{\mu\nu}(x) = const$
and $g_{\textrm{a}} = 0$, the eigenstate problem can be solved in much the same way as
described in \eqref{e33} and \eqref{e34}. Indeed, the transformation analogous to
\eqref{e33} reads as follows:
\begin{eqnarray}
   \label{e36}
   \psi^{(g)}_{n}(\ev{r}) &=& \exp\left\{-\frac{i g}{\hbar c}\ev{r}
                  \left( [\ev{v}\ev{E}] - v^0\ev{H} \right)\right\}\psi^{(0)}_n(\ev{r}),\\
   E_n^{(g)} &=& E^{(0)}_n + \ev{v}\cdot\ev{H}.
\end{eqnarray}
Again, the spectrum is shifted by a constant value, though depending on the direction of
the magnetic field.
%
However, the change in the wavefunctions could possibly affect, for instance, the
properties of synchrotron radiation in a homogeneous magnetic field.

Nonetheless, we confine ourselves to demonstrating the prospects of unitary
transformations for solving wave equations containing Lorentz-violating terms. In the
following sections, similar techniques will be used to obtain the solutions in the case
of the $b_\mu$ coupling, and to study the dynamics of a bound electron in such a
background.
\section{Hydrogen-like atom. Quasi-relativistic approach}\label{sec4}
Consider first the Pauli Hamiltonian \eqref{eq:hPauli2} within the first order in
$b_\mu$:
\begin{equation}
  \hat{h}_P = \frac{\hat{\ev\Pi}^2}{2 m_e} + e A_0
  - \frac{e\hbar}{2m_ec}\ev{\sigma}\ev{H} + \ev{\sigma}\ev{b}.
\end{equation}
We suppose that $A_\mu(x)$ is taken in the Coulomb gauge with
\begin{eqnarray}
  \frac{\pd A_0}{\pd t} &=& 0,\\
  \divg\ev{A} &=& 0.
\end{eqnarray}
Make an inverse gauge-like shift of the momentum ($\hat{\ev{\Pi}} \to \hat{\ev{P}}$)
performing a \textit{unitary} transformation:
\begin{eqnarray}
    \Phi_P &\rightarrow& \Phi'_P = \hat{U}_P\Phi_P, \quad
    \hat{h}_P \rightarrow \hat{h}'_P = \hat{U}_P\hat{h}_P\hat{U}_P\hc,\qquad
    \hat{U}_P\equiv \exp\left\{-\frac{i b_t}{\hbar} \ev{\sigma}\cdot\ev{r}\right\};\\
    \hat{h}'_P &=& \frac{\hat{\ev{P}}^2}{2m_e}+ e A_0
     - \left(\frac{e\hbar}{2m_ec} \ev\sigma + \hat{\ev{\mu}}_A\right)\ev{H} + \ev{\sigma}\cdot\ev{b},  \label{U_P}\\
    \hat{\ev\mu}_A &=& \frac{e b_t}{m_e c}[\ev{\sigma}\ev{r}]. \label{mu_A}
\end{eqnarray}
It is clear that the transformation reduces the Lorentz-violating
interaction to a modification of the electron magnetic moment, which
acquires a \CPT{}-odd correction $\hat{\ev{\mu}}_A$. Consequently,
the terms of the first order in $b_t$ vanish in the transformed
Hamiltonian as the external magnetic field $\ev{H}$ is turned off.
In particular, within the approximation used, the eigenstate problem
in an \textit{electric} field would look quite conventional after
the transformation. In a relativistic theory discussed in section
\ref{page:MuaDa}, an \textit{electric} dipole moment correction also
arises but it vanishes in the nonrelativistic approximation.

Let $A^\mu = \{ \phi(r), \ev{0}\}$ and $\ev{b} = \{0, 0, b_z\}$, $\phi(r)$ being the potential of the nucleus
initially considered as spherically-symmetric, but not mandatory the Coulomb potential.\footnote{Indeed, due to
radiative corrections, Coulomb attraction becomes stronger than $\sim 1/r$ at short distances contributing in
the Lamb shift of electron eigenstates \cite{LambShift}. The Lamb shift however originates from the three
one-loop corrections to electrodynamics including electron mass renormalization, its anomalous magnetic moment
and the modification of the Coulomb law. For $s$-states, the first of them makes a major contribution to the
Lamb shift.}
The problem resembles that of an electron in a homogeneous magnetic field $\ev{H}_b$ but for the only
difference: now there is no coupling in the kinetic term (that is, we have $\hat{\ev{p}}$ for the momentum
instead of $\hat{\ev{p}}-\frac{e}{c}\ev{A}_b$, where $\ev{H}_b = \rot\ev{A}_b$). The coupling to the external
field $\ev{b}$ involves only the spin degrees of freedom but not the orbital ones. The energy eigenstates can be
easily obtained in the transformed representation and then transformed back to the initial one:
\begin{eqnarray} \label{e31}
  (\Phi'_P)_{n l m_l m_s}(\ev{r}) &=& R_{nl}(r)Y_{l,m_l}(\ev{r}/r) \chi_{m_s};\\
    (\Phi_P)_{n l m_l m_s}(\ev{r}) &=& R_{nl}(r)Y_{l,m_l}(\ev{r}/r)
    \left(1+\frac{i b_t}{\hbar}\ev{\sigma}\cdot\ev{r}\right)\chi_{m_s};\\
    E_{n l m_s} &=& E^{(0)}_{n l} + 2 b_z m_s,
\end{eqnarray}
where  $n=1,2,3,\ldots$, $l = \overline{0, n-1}$, $m_l =
\overline{-l,l}$, $m_s = \pm 1/2$ are the quantum numbers denoted
according to a common convention; $\chi_{m_s}$ are the spin
$z$-component eigenvectors. $R_{nl}(r)$ and $E_{nl}^{(0)}$ are the
radial wavefunction and the energy in the $b_0 = 0$ case,
respectively. In the Coulomb case, we have \cite{LL3}
\begin{eqnarray}\label{eq:R_nl}
    R_{nl}(r) &=& \frac{2 Z^{3/2}}{n^2r_{\textrm{B}}^{3/2}}\sqrt{\frac{(n-l-1)!}{(n+l)!}}
                 e^{-\rho/2}\rho^l L_{n-l-1}^{(2l+1)}(\rho),\\
    E^{(0)}_{n l} \to E^{(0)}_n &=& -\frac{Z^2\hbar R}{n^2},
\end{eqnarray}
where $\rho = 2 Z r / n r_{\textrm{B}}$, $r_{\textrm{B}} = \hbar^2/m_ee^2$ is the Bohr radius, $R = m_e e^4 /
2\hbar^3$ is the Rydberg constant, and $L^{(\nu)}_{k}$ denote the generalized Laguerre polynomials:
\begin{equation}\label{eq:Laguerre}
  L^{(\nu)}_k(\rho) = \frac{1}{k!} \rho^{-\nu} e^{\rho} \frac{d^n}{d\rho^n}
                                   \left(\rho^{\nu+n} e^{-\rho}\right), \quad \RRe{\nu} > 0, \quad k =
                                   0,1,2,\ldots
\end{equation}

The solution obtained shows that, with respect to the transformed representation, the only effect the presence
of $\ev{b}$ generates \textit{in the leading order} is a removed degeneracy over spin quantum number $m_s$, with
the energy splitting being $\lesssim10^{-4}\mbox{Hz}$. Neither the spectrum nor the eigenfunctions are affected
by $b_0$, \textit{only} the interaction with the external magnetic field is.

The $\ev{b}$-induced energy splitting into a doublet is a formal result of solving the
eigenstate problem in a $1/c$-approximation that does not hold true when the spin-orbit
interaction is considered that removes the degeneracy over quantum number $j$. The
correct splitting magnitude can be estimated by means of a perturbation theory. In the
absence of $\ev{b}$, the spectrum remains degenerate over $l$ and $m_j$. The action of
the perturbation term $\ev\sigma\cdot\ev{b}$, however, preserves these quantum numbers,
so the perturbation theory can be applied to the atom as to a non-degenerate system. This
situation is typical for the anomalous Zeeman effect \cite{STZ, LL3}.

For the $\ev\sigma\ev{b}$ term, the energy correction was first estimated in
\cite{Brazil:SMEDirac} but we shall do it once again. First, let $b^\mu = \{0,0,0,b_z\}$.
Following the arguments explained in the preceding paragraph, take $\cket{n l j m_j}$ for
the eigenstates in the $\ev{b} = 0$ case. Using the general expressions for them
\cite{LL4, STZ},
\begin{eqnarray}
    \left<\ev{r}|n l j m_j\right> &=& R_{n l j}(r) Y^l_{j m_j}(\ev{r}/r), \label{eq:R_nlj}\\
    Y^l_{j m_j} &=& \frac{1}{\sqrt{2l+1}}
    \okcol{\sqrt{l+1/2+\varkappa m_j}}{\varkappa \sqrt{l+1/2-\varkappa m_j }},\\
    \varkappa &\equiv& (-1)^{(l-l'+1)/2} = \pm1 \text{ for } j = l \pm 1/2,
    \qquad l' \equiv 2j - l, \label{eq:varkappa&l'}
\end{eqnarray}
we obtain:
\begin{eqnarray}
    \Delta E^{(\ev{b})}_{n l j m_j} =
    \brac{n l j m_j} \ev\sigma \ev{b} \cket{n l j m_j} &=&
    \int\limits_0^\infty R^2_{n l j}(r) r^2dr \cdot
    \frac{b_z}{2l+1} \left((l + 1/2 + \varkappa m_j) - (l + 1/2 - \varkappa m_j) \right)
    =\nonumber\\&=&
    \int\limits_0^\infty R^2_{n l j}(r) r^2dr \cdot \frac{2 \varkappa m_j }{2l+1} b_z
    = \frac{2 \varkappa m_j }{2l+1} b_z,
\end{eqnarray}
that is, twice the result obtained in \cite{Brazil:SMEDirac}. Since the corrections
induced by $\ev{b}$ are minuscule, we will further treat $b^\mu$ as a purely  timelike
4-vector, with the time component $b_0$.

In search for $b_0$-corrections to the eigenstates, we shall resort to the
$1/c^2$-approximation in the eigenstate problem. Consider the Coulomb case with $e\phi(r)
= -Ze^2/r$ within the first-order approximation in $b^\mu = \{c b_t, \ev{0}\}$. A
spectacular feature of this case is that the solutions can be explicitly expressed via
their conventional form (for $b_0 = 0$). The correspondence is generated again with a
unitary transformation:
\begin{eqnarray}
  \hat{h} &=& \hat{U}\hc\hat{h}|_{b_0 = 0}\hat{U}, \qquad
  \hat{U} = \exp\left\{-\frac{i b_t}{\hbar}\left(1 + \frac{Ze^2}{2m_ec^2r}\right)
                          \ev{\sigma}\cdot\ev{r} \right\},\\
  \hat{h} &=& \frac{\hat{\ev\pi}^2}{2m_e}
            \left(1 - \frac{\hat{\ev\pi}^2}{4m_e^2c^2}\right) - \frac{Z e^2}{r}
            + \frac{Ze^2\hbar^2}{4m_e^2c^2}
            \left(\frac{\ev{\sigma}\hat{\ev{l}}}{r^3}+2\pi\delta(\ev{r})\right).
\end{eqnarray}
As a result we obtain
\begin{eqnarray}
  \Phi_{nljm_j}(\ev{r}) &=& R_{nlj}(r)\left\{Y^{l}_{jm_j}(\ev{r}/r)
  - \frac{\varkappa b_t r}{\hbar}
  \left(1 + \frac{Ze^2}{2m_e c^2 r}\right)Y^{l'}_{jm_j}(\ev{r}/r)\right\},\\
  E &=& E^{(0)}_{nj} = -\frac{Z^2 \hbar R}{n^2}\left[1+
       \frac{Z^2\alpha^2}{n}\left(\frac{1}{j+1/2} -\frac{3}{4n}\right)\right],
\end{eqnarray}
where $\varkappa$ and $l'$ are defined in \eqref{eq:varkappa&l'}. The radial functions
$R_{nlj}(r)$ remain the same as in the $b_0 = 0$ case (see \eqref{eq:R_nlj}). In the
nonrelativistic limit, they take the form \eqref{eq:R_nl}.

Thus, no corrections to the energy spectrum are present due to $b_0$, within the
$1/c^2$-approximation. Further analysis will show that there are no corrections of the
first order in $b_0$ (see section \ref{sec5}). Nevertheless, the perturbative method used
in \cite{Brazil:SMEDirac} to retrieve the energy corrections due to the term
$-\frac{b_0}{m_e c}\ev\sigma\cdot\hat{\ev{p}}$ is incorrect. The spectrum is degenerate
over $l$ if $b_0 = 0$, while the perturbation operator is $\Parity$-odd, and hence
changes the $l$ quantum number. The expectation value of such an operator clearly
vanishes in a state possessing a definite $l$, and thus definite parity $P = (-1)^l$. The
energy shift may not vanish, however, for some superposition of the states with opposite
parities. This is common for the \textit{linear Stark effect} \cite{LL3, STZ} that occurs
due to a degeneracy of the hydrogen spectrum. The perturbation theory for a degenerate
system must be employed instead of a simple averaging.
Despite the above remarks, the methods employed in \cite{Brazil:SMEDirac}, have led to correct results.

 In conclusion, we will show how the correction to the magnetic moment may cause an
appearance of a nonzero \textit{anapole} moment of the atomic \textit{orbital}
\cite{Zeldovich:Anapole}. This is a classical quantity ascribed to a
\textit{parity-nonconserving} system (such systems exist in the conventional Standard
Model due to weak interaction \cite{BorZhukTern:Anapole}) adding an interaction term of
the form $-\ev{T}_{\text{Z}}\cdot\rot{\ev{H}}$ to the Hamiltonian of the system, with
$\ev{T}_{\text{Z}}$ being the anapole moment. Consider a hydrogen atom in the ground
state $1s_{1/2,{m_j}}$ where lower indices indicate the electron total angular momentum
and its $z$-projection. Averaging the \CPT-odd term $-\hat{\ev{\mu}}_A\cdot\ev{H}$ in
this state yields:
\begin{equation}
  V_{\text{Z}} \equiv \avg{-\hat{\ev{\mu}}_A\cdot\ev{H}(\hat{\ev{r}})} =
  -\avg{\hat{\ev{\mu}}_A\left(\ev{H}(\ev0) +
  (\hat{\ev{r}}\cdot\ev\nabla)\ev{H}(\ev0)+\ldots\right)},
\end{equation}
where $\ev{r} = \ev0$ points to the center of the Coulomb field. The ground $1s$ state
possesses a definite parity $(-1)^l = +1$ and, in addition, a spherical symmetry, hence
\begin{eqnarray}
  \avg{\hat{\ev{\mu}}_A} &=& 0,\\
  \avg{\hat{x}_i\hat{x}_k} &=& \frac13\delta_{ik}\avg{\hat{r}^2}, \qquad i,k = 1,2,3.
\end{eqnarray}
With the help of expressions \eqref{e31} and \eqref{eq:R_nl}, one can easily find that
\begin{equation}
 \avg{\sigma_i \hat{x}_k \hat{x}_n} = 2 r_{\textrm{B}}^2 \delta_{i 3} \delta_{kn}\cdot {m_j},
  \qquad i, k, n = 1, 2, 3.
\end{equation}
which results in the following:
\begin{eqnarray}
  V_{\text{Z}} &\approx&
  -\frac{2 e b_0 r_{\textrm{B}}^2}{m_e c^2} {m_j} \epsilon_{3 i k}\pd_i H_k =
                       -\ev{T}_{\text{Z}}\cdot\rot\ev{H},\\
  \ev{T}_{\text{Z}} &=& 2 e r_{\textrm{B}}^2 \left(\frac{b_0}{m_e c^2}\right) {m_j} \ev{e}_3,
\end{eqnarray}
where $\ev{e}_3$ is the basis unit vector along the $z$-axis.
\section{Series expansion of the Dirac equation with respect to $b_0$}\label{sec5}
In this section we discuss the case $b^\mu = \{b^0, \ev{0}\}$ and
$A^\mu = \{A^\ext_0(x) + \phi(r), \ev{A}^\ext(x)\}$ using the
Heaviside units, with $\hbar = c = 1$, $\alpha = \frac{e^2}{4\pi}$.
Consider the Hamiltonian \eqref{H} and transform the corresponding
wave equation using the \textit{gauge-invariant unitary}
transformation:
\begin{eqnarray}
  \psi(x) &\rightarrow& \tilde\psi(x) = e^{-ib_0\hat\Delta_A}\psi(x),\\
  \hat{H} - i \ddt &\rightarrow& \hat{\tilde{H}} - i \ddt =
  e^{-ib_0\hat\Delta_A}\left(\hat{H} - i\ddt\right) e^{ib_0\hat\Delta_A};\\
  \hat{\Delta}_A &=& \ev{\Sigma}\cdot\hat{\ev{r}} - \frac{i}{m_e}(\ev{\Sigma}\cdot\hat{\ev{L}} +
  1)\gamma^0\gamma_5,\\
  \hat{\ev{L}} &=& [\hat{\ev{r}}\hat{\ev{P}}] = - [\hat{\ev{P}}\hat{\ev{r}}].
\end{eqnarray}
Restricting ourselves to the second-order approximation in $b_0$, we
obtain:
  \begin{equation}\label{eq:hTilde}
     \hat{\tilde{H}} \approx
     \ev\alpha(\hat{\ev{p}} - e \ev{A}^\ext) + \beta m_e + e(\phi + A_0^\ext)
     -\frac{b_0^2}{m_e}\hat{f}\gamma^0
     - \hat{\ev{d}}_A \ev{E}^\ext - \hat{\ev\mu}_A \ev{H}^\ext
     + H^{(2)}_\Int[A^\ext],
   \end{equation}
with $\hat{f} \equiv \ev{\Sigma}\hat{\ev{l}} + 1$. $H^{(2)}_\Int[A^\ext]$ stands for the
second-order terms in $b_0$ describing the interaction with the external field
$A_\mu^\ext$. \label{page:MuaDa} Additional electric and magnetic dipole moment operators
read as follows:
\begin{eqnarray}
    \hat{\ev\mu}_A &=& \frac{eb_0}{m_e}\gamma^0[\ev{\Sigma}\ev{r}],\\
    \hat{\ev{d}}_A &=& -i\gamma_5\hat{\ev\mu}_A = -\frac{i e b_0}{m_e}[\ev\gamma\ev{r}].
\end{eqnarray}
As we can see, no non-linear terms in the external field are present up to the first
order in $b_0$, inclusively. Moreover, the moment $\hat{\ev{d}}_A$ couples with the
\textit{external} field only but \textit{not} with the spherically-symmetric `background'
field $\phi(r)$, because for such a field о $\hat{\ev{d}}_A\cdot(-\ev\nabla\phi) = 0$.
The same situation holds in every higher order of the expansion due to the fact that
$[\hat{\Delta}_A, \phi(r)] = 0$. For the same reason, the expressions for operators
$\hat{\ev{d}}_A$ and $\hat{\ev\mu}_A$ are not affected by $\phi(r)$, in particular, they
maintain their form for a \textit{free} electron. Our approach however is applicable only
to systems with the effective size much less than $1/b_0 \gtrsim 10^{-3}\text{cm}$.

We did not obtain a \CPT-odd correction coupling to the electric field in the
$1/c$-approximation because, in contrast to $\hat{\ev{\mu}}_A$, $\hat{\ev{d}}_A$ is a
block-off-diagonal operator,
\begin{equation}
\hat{\ev{d}}_A = -\frac{i e b_t}{m_e c}
  \okmatr{0}{\cross{\ev\sigma}{\ev{x}}}{-\cross{\ev\sigma}{\ev{x}}}{0},
\end{equation}
which mixes the `upper' and the `lower' 2-component spinors of the wavefunction. The
`lower' spinor vanishes in the non-relativistic limit (see eq. \eqref{e26}), and so does
the operator $\hat{\ev{d}}_A$. Instead, $\hat{\ev{\mu}}_A$ is a block-diagonal matrix
that mixes the `upper' spinors with themselves and consequently it does not vanish in the
nonrelativistic limit.

Let $A_\mu^\ext = 0$, then the spherical symmetry allows us to search for the
eigenfunctions in the form
\begin{equation}\label{psiAnsatz}
     \tilde{\psi}_{n_rljm_j}(\ev{r}, t) = \okcol{R^{(u)}_{n_rlj}(r) Y^l_{jm_j}(\ev{r}/r)}
                                          {\varkappa R^{(v)}_{n_rlj}(r) Y^{l'}_{jm_j}(\ev{r}/r)},
\end{equation}
where $n_r \equiv n - j - 1/2$ is the radial quantum number and $n$ is the principal
quantum number, and $l$, as usual in the relativistic theory, determines the parity of
the state $P = (-1)^l$, but not its orbital momentum. Operators $\hat{\ev{d}}_A$ and
$\hat{\ev\mu}_A$ have vanishing expectation values in such a state.
$\tilde\psi$ is the eigenfunction of the operator $\hat{f}\gamma^0$, with the eigenvalue
$f \equiv \varkappa(j+1/2)$. In the case under consideration, the transformed Hamiltonian
\eqref{eq:hTilde} is the sum of its conventional value (for $b_0 = 0$) and a term
proportional to $\hat{f}\gamma^0$; $\tilde\psi$ is an eigenfunction for both of them if
the radial functions $R^{(u,v)}$ are taken the same as those for the $b_0 = 0$ case. The
energy value which responds to $\tilde\psi$ is
\begin{equation}
     E = \tilde{E} = E^{(0)}_{n_r l j} - \varkappa(j+1/2)\frac{ b_0^2}{m_e}
               = E^{(0)}_{n_r l j} \pm (j+1/2)\frac{ b_0^2}{m_e}
               \quad\mbox{ for } l = j\pm 1/2.
\end{equation}
An additional $b_0$-induced second-order energy splitting therefore arises:
\begin{equation}\label{dE}
   \Delta{E}(j) \equiv E_{n_r,j+1/2,\,j} - E_{n_r,j-1/2,\,j} = (2j+1)\frac{b_0^2}{m_e}.
\end{equation}
This term originates from parity violation due to the $b_\mu$-induced violation of \CPT{}
and removes the degeneracy over $l$ in the Coulomb field case. $\abs{\Delta{E}(j)}
\lesssim 10^5\text{Hz}$ for $j = 1/2$, that is, four orders of magnitude smaller than the
Lamb shift \cite{LambShift}. Nonetheless, in contrast to the latter one existing mainly
for $s$-states, the splitting \eqref{dE} \textit{increases} with growing $j$.

The eigenfunctions in the initial representation are obtained after
performing the inverse transformation:
\begin{equation}
     \psi_{n_rljm_j}(\ev{r}) = e^{-b_0^2 f^2/2m_e^2} e^{-b_0^2 r^2/2}
       \okcol{R^{(u)} Y^{l}_{jm_j} + b_0\varkappa \left(\frac{f}{m_e}R^{(v)} - r R^{(u)}\right)
          Y^{l'}_{jm_j}}
          {\varkappa R^{(v)} Y^{l'}_{jm_j} + b_0\left(\frac{f}{m_e}R^{(u)} + r R^{(v)}\right)
          Y^{l}_{jm_j}}.
\end{equation}
The presence of the admixture of spherical spinors with the
different value of the orbital quantum number ($l'$) breaks the
parity of the states. `Probability distribution' $\psi\hc\psi$ is
not affected however, compared with the conventional ($b_0 = 0$)
case, within the chosen approximation. Since the conventional
solution in the Coulomb case is well-known \cite{LL4}, we can
explicitly find the second-order approximation for the
eigenfunctions in the $b_0 \ne 0$ Coulomb case:
\begin{eqnarray}
     E^{(0)}_{n_rj} &=& m_e\left(1 + \left(\frac{Z\alpha}{\gamma+n_r}\right)^2\right)^{-1/2},\\
     \left.\begin{array}{c}
       R^{(u)}\\
       R^{(v)}
     \end{array}\right\}
         &=&
     \pm (2\lambda)^{3/2}
     \left( \frac{   (m_e\pm E^{(0)}_{n_r j}) n_r!    }
                 {4m_e   \frac{Z\alpha m_e}{\lambda}
                         \left(f + \frac{Z\alpha m_e}{\lambda}\right)
                         \Gamma(2\gamma + n_r + 1)
                               }
     \right)^{1/2}
     e^{-\lambda r}   (2\lambda r)^{\gamma-1}  \times \nonumber\\
     &&\times\left(
                \left(f + \frac{Z\alpha m_e}{\lambda}\right)
                      L^{(2\gamma)}_{n_r}(2\lambda r)
                \pm (1 - \delta_{n_r,\,0}) (2\gamma + n_r)
                      L^{(2\gamma)}_{n_r - 1}(2\lambda r)
             \right),
      \\
      \lambda &\equiv& \sqrt{m_e^2 - {E^{(0)}_{n_rj}}^2},\qquad
      \gamma = \sqrt{(j+1/2)^2 - (Z\alpha)^2},
\end{eqnarray}
with $L^{(2\gamma)}_{n_r}$ being the generalized Laguerre polynomials defined in \eqref{eq:Laguerre}.The
expression for the energy demonstrates an alternative mechanism of removing the degeneracy over $l$, different
from that connected with the one-loop corrections in quantum electrodynamics.
\section{Specific radiative properties of a hydrogen atom induced by $b_0$}
\label{sec6} Finally, we demonstrate an example with \CPT{} and Lorentz violation leading
to radiative effects specific for the $b_0 \ne 0$ case and \textit{linear} in $b_0$.
Following the system of units convention used in section \ref{sec3}, we assume $b^\mu =
\{c b_t, \ev{0}\}$. Since the primary goal of this section is to obtain the leading-order
$b_0$-induced terms in the radiation distribution, we restrict ourselves to the Pauli
approximation and consider the radiation of a hydrogen ($Z = 1$) atom. Upon the
transformation \eqref{U_P}, the only term remaining with $b_0$ is
$-\hat{\ev\mu}_A\cdot\ev{H}$. This term violates the spatial parity of the atom. With the
use of the standard formulas \cite{STZ}, we find the angular distribution of spontaneous
radiation probability:
\begin{eqnarray}
      \frac{dW_{fi}(\ev{k},\lambda)}{d\Omega_{\ev{k}}} &=&
      \frac{\omega^3}{2\pi\hbar c^3}
      \left|\ev{e}^{(\lambda)*}(\ev{k})\cdot \ev{\mathfrak{m}}_{fi}(\ev{k})\right|^2,\\
      |\ev{k}| &=& \omega/c = (E_i - E_f)/\hbar c > 0, \qquad \lambda = 1,2;\\
      \hat{\ev{\mathfrak{m}}} &=& e \hat{\ev{r}}
      - \frac{ie}{2}(\ev{k}\cdot\hat{\ev{r}})\hat{\ev{r}}
      - \left[\frac{\ev{k}}{k}\times\hat{\ev\mu}\right],\\
      \hat{\ev{\mu}} &=& \frac{e\hbar}{2m_e c} (\hat{\ev{l}} + \ev\sigma) +
      \hat{\ev\mu}_A,
\end{eqnarray}
where $\ev{k}, \lambda$ are the photon momentum and polarization, and $\ev{e}^{(\lambda)}(\ev{k})$ is the
polarization vector. $\brac{f}$ and $\cket{i}$ denote the final and the initial electron states. The correction
$\hat{\ev{\mu}}_A$ to the magnetic moment operator $\hat{\ev\mu}$ is defined in \eqref{mu_A}.

\inspic{Angular distribution of spontaneous radiation for $2p_{1/2,1/2}\to1s_{1/2,-1/2}$ transition}
       {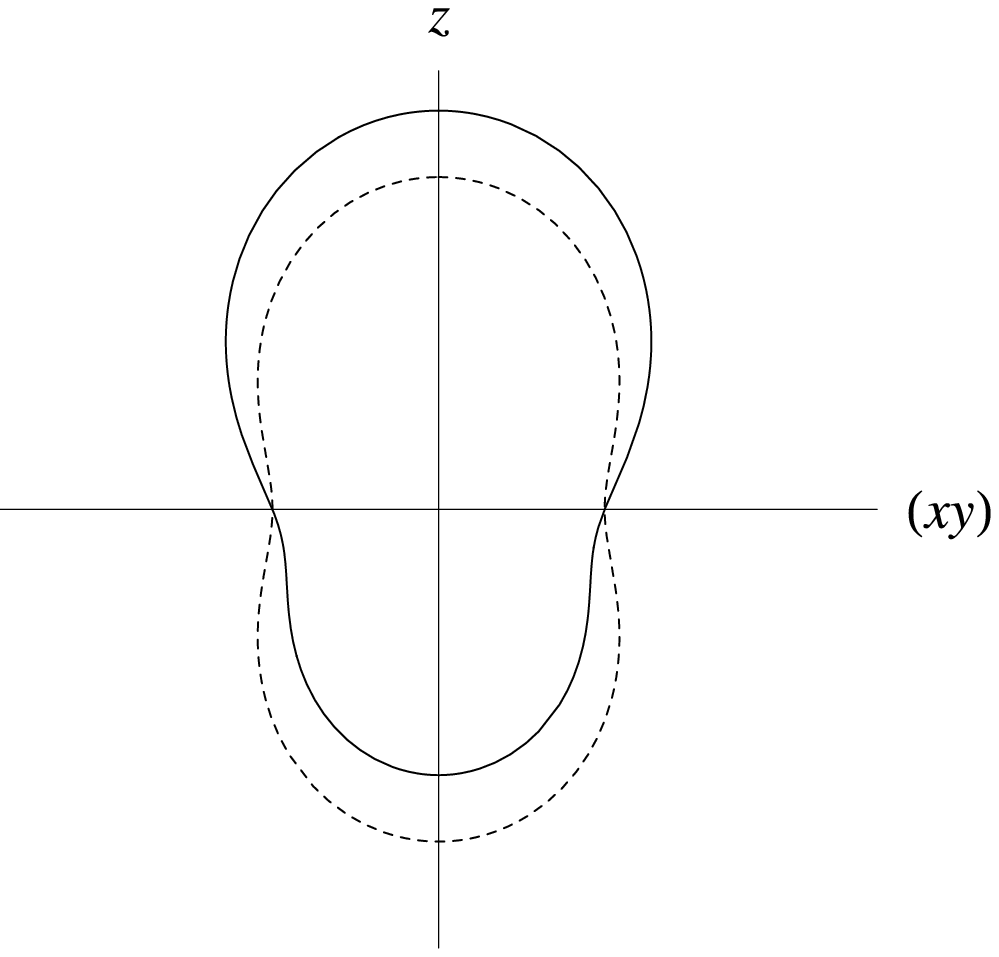}
       {0.355\textwidth}
Radiation processes allowed due to the parity-violating interaction with $\hat{\ev{\mu}}_A$ (further referred to
as $Aj$-radiation with $j$ denoting the photon angular moment) are restricted by the same selection rules as
those for the $E1$-radiation, the corresponding matrix element having the form typical for $M1$-radiation. Thus
$A1$- and $E1$-photons have the same multipolity but the opposite parity. Linear in $b_0$ corrections to the
angular distribution occur due to the interference between the $E1$- and the $A1$-radiation. Consequently, they
vanish over the whole sphere because of the cancellation of spherical spinors with different parities, resulting
in no linear in $b_0$ terms present in the \textit{total} transition rate.

The interference term does not vanish, for example, for the transition
$2p_{1/2,1/2}\to1s_{1/2,-1/2}$. The calculations show that after averaging over the
photon polarizations $\lambda = 1,2$, the resulting angular distribution reads ($\Theta$
is the angle between $\ev{k}$ and the $z$-axis):
\begin{equation}
  \frac{dW}{d\Omega_{\ev{k}}} =
   \frac{512\alpha^3 R}{6561\pi}\left\{1 + \cos^2\Theta
                                + \frac{8 b_0}{m_e c^2}\cos\Theta\right\}. \label{w}
\end{equation}
As we can see, the presence of $b_0$ induces the violation of the conventional
`$\ev{k}$-parity' of the distribution (the radiation rates in the opposite directions
differ in the $b_0 \ne 0$ case). The relative magnitude of this violation is of the order
$|b_0|/m_e c^2 \lesssim 2\cdot 10^{-8}$. Distribution \eqref{w} is depicted in
fig.\ref{pic:fig1}, with the dotted curve related to the $b_0 = 0$ case. To make the
picture more vivid, we chose $b_0/m_ec^2 = 0.05$.
\par For unpolarized atoms, i.e. after averaging over ${m_j}$, ${m_j}'$ quantum numbers, the
spherical symmetry is restored in the distribution, with no linear
in $b_0$ $\ev{k}$-odd contributions present. This is the consequence
of $SO(3)$-invariance unbroken even in the $b_0 \ne 0$ case (while
$O(3)$ symmetry is broken since $b_0$ is a pseudoscalar).

We left aside the problem of polarization of atoms. If one uses
Zeeman effect in a homogeneous magnetic field to obtain the
polarization, then this magnetic field would also lead to parity
violation due to the interaction with $\hat{\ev{\mu}}_A$. Another
way the external magnetic field can break the atomic \Parity-parity
is that in the reference frame of a moving atom, an additional
electric field will be induced that breaks the parity. The
distribution of radiation of moving atoms can also be shifted due to
aberration. These problems need further consideration. In the
present paper however, we just demonstrated yet another scenario of
$\Parity$-parity violation in atomic transitions.
\section{Conclusion}\label{sec7}
In this paper, we have considered several solutions of the Dirac
equation in the framework of the Standard Model Extension with
particular types of Lorentz violation.
The $1/c^2$-approximation for the extended Dirac equation was derived in the background
of the axial vector \SME{}-coupling $b_\mu$. The expansion of the \textit{relativistic}
Dirac equation with respect to $b_0$ has been employed to solve the eigenstate problem
for an electron in a spherically-symmetric potential well. The unitary transformation was
found that was used to express the solutions with $b_0 \ne 0$ in terms of solutions for
$b_0 = 0$, with the second order accuracy with respect to $b_0$. Explicit solutions have
been obtained in the case of the Coulomb potential, demonstrating a specific
$b_0$-quadratic energy splitting. The degeneracy over the orbital quantum number is
removed, and it was shown that the corresponding energy splitting does not vanish for
large $j$.

In addition, unitary transformations were used to obtain the \textit{exact} eigenstates
in the case of the coupling $-g \bar\psi\gamma^\mu v^\nu F^*_{\mu\nu}\psi$ with constant
homogeneous electromagnetic field $F_{\mu\nu}$ \cite{Belich}.

The unitary transformation made it possible to obtain the $b_0$-corrections to the
operators of $E1$ and $M1$ moments of the electron. These moments effectively lead to an
existence of the \textit{anapole} moment of the orbital \cite{Zeldovich:Anapole}.

Finally, the distribution of spontaneous radiation of a polarized hydrogen atom was shown
to lose its central symmetry in the $b_0 \ne 0$ case, due to the violation of spatial
parity. The results obtained can be treated only as an illustration of the application of
the model adopted. There are other physical effects that should also be considered
together with the one discussed in this paper.
\section{Acknowledgements}
The authors are grateful to A.V.Borisov, D.Ebert, and A.V.Lobanov for helpful
discussions.

\end{document}